\documentclass[aps,prd,reprint,preprintnumbers,groupedaddress,nofootinbib]{revtex4-1}

\usepackage{xspace}
\usepackage{graphicx}
\usepackage{amsmath}

\usepackage{url}

\newcommand{\pythia}[1]{\textsc{Pythia\xspace #1}}

\newcommand{\fastjet}[1]{\textsc{FastJet\xspace #1}}

\newcommand{\akt}{anti-$k_T$}
\newcommand{\zcut}{z_\text{cut}}

\DeclareRobustCommand{\Sec}[1]{Sec.~\ref{#1}}

\DeclareRobustCommand{\App}[1]{App.~\ref{#1}}

\DeclareRobustCommand{\Fig}[1]{Fig.~\ref{#1}}

\DeclareRobustCommand{\Eq}[1]{Eq.~(\ref{#1})}

\DeclareRobustCommand{\Ref}[1]{Ref.~\cite{#1}}
\DeclareRobustCommand{\Refs}[1]{Refs.~\cite{#1}}

\DeclareMathOperator*{\argmin}{argmin}

\begin{document}

\preprint{MIT-CTP 4563}

\title{Aspects of Jets at 100 TeV}

\author{Andrew J. Larkoski}
\author{Jesse Thaler}
\affiliation{Center for Theoretical Physics, Massachusetts Institute of Technology, Cambridge, Massachusetts 02139, USA}


\begin{abstract}

We present three case studies at a 100 TeV proton collider for how jet analyses can be improved using new jet (sub)structure techniques.  First, we use the winner-take-all recombination scheme to define a recoil-free jet axis that is robust against pileup.  Second, we show that soft drop declustering is an effective jet grooming procedure that respects the approximate scale invariance of QCD.  Finally, we highlight a potential standard candle for jet calibration using the soft-dropped energy loss.  This latter observable is remarkably insensitive to the scale and flavor of the jet, a feature that arises because it is infrared/collinear unsafe, but Sudakov safe. 

\end{abstract}


\maketitle

\section{Introduction}
\label{sec:int}

The LHC has ushered in a new era of precision jet physics, with advances in both QCD calculations and jet analysis techniques.  By now, jets at the LHC are robust objects, characterized not just by their overall energy and direction but also by their substructure \cite{Abdesselam:2010pt,Altheimer:2012mn,Altheimer:2013yza}.  On the longterm horizon is a future hadron collider at 100 TeV \cite{Gershtein:2013iqa,cern100,china100}, which will open a new kinematic regime at high energies.  Given the jet successes at the LHC, it is worth studying whether new jet techniques might improve the physics capabilities of a 100 TeV machine.

In this paper, we highlight three potentially powerful ways to define and study jets:  winner-take-all (WTA) axes \cite{Bertolini:2013iqa,Larkoski:2014uqa,Salambroadening}, soft drop declustering \cite{Larkoski:2014wba}, and Sudakov-safe observables \cite{Larkoski:2013paa}.  Most of these methods have been introduced elsewhere, along with detailed analytical calculations.  The goal here is to demonstrate the utility of these procedures in a high-energy and high-luminosity environment.  Our focus is on jets at 100 TeV, but of course, these same studies are relevant for Run II of the LHC at 14 TeV.

Leading up to the LHC era, one of the key concepts (if not \emph{the} key concept) in jet physics was ``infrared and collinear (IRC) safety'' \cite{Ellis:1991qj}, which characterizes whether an observable can be predicted at fixed order in perturbative QCD.  Looking towards a 100 TeV machine, we think that two important concepts in jet physics will be ``recoil insensitivity'' and ``Sudakov safety''.
\begin{itemize}
\item \textit{Recoil insensitivity}.  Jets can be affected by various kinds of contamination from uncorrelated radiation, including perturbative soft radiation and non-perturbative underlying event activity.  As one goes to higher energies and luminosities, pileup (multiple proton-proton collisions per bunch crossing) becomes an increasingly problematic source of jet contamination.  Just by momentum conservation, any uncorrelated radiation can significantly displace the momentum axis of a jet from the direction associated with the initiating hard parton.  This effect is known as ``recoil'' \cite{Catani:1992jc,Dokshitzer:1998kz,Banfi:2004yd,Larkoski:2013eya}, and can be present even for IRC-safe observables.  For this reason, one would like to work with ``recoil-free'' observables which are insensitive to this effect.
\item \textit{Sudakov safety}.  IRC safety is a necessary condition for observables to be computable order by order in a perturbative $\alpha_s$ expansion.  Recently it was realized that certain IRC unsafe observables can still be calculated using techniques from perturbative QCD.  In particular, ``Sudakov safe'' observables \cite{Larkoski:2013paa} cannot be expressed as a Taylor series in $\alpha_s$, but they can still be calculated perturbatively by using resummation to capture all-orders behavior in $\alpha_s$.  Given their different analytic structures, IRC safe and Sudakov safe observables can be sensitive to very different physics.  
\end{itemize}

To highlight these concepts, we present three case studies of jets at 100 TeV.  In \Sec{sec:pu}, we discuss the recoil sensitivity of the standard jet axis, and show that the WTA recombination scheme \cite{Bertolini:2013iqa,Larkoski:2014uqa,Salambroadening} results in jets whose axis is recoil insensitive and hence robust to high levels of pileup.  In \Sec{sec:sd}, we show that the soft drop declustering procedure \cite{Larkoski:2014wba} is a powerful technique for pileup mitigation,  and we exhibit its performance for dijet resonance reconstruction.  In \Sec{sec:sd_eloss}, we present a Sudakov-safe \cite{Larkoski:2013paa} and quasi-conformal observable defined via the soft drop procedure, which is only weakly dependent on the value of $\alpha_s$ (and as such has weak energy-scale dependence).  This observable is also remarkably similar between quark and gluon jets, providing a potential ``standard candle'' for 100 TeV jet calibration.

The studies in this paper are based on Monte Carlo simulations of 100 TeV proton-proton collisions.  All event generation, parton showering, and hadronization is done with \pythia{8.183} \cite{Sjostrand:2006za,Sjostrand:2007gs} at Born-level only with no fixed-order corrections.  Jet analyses are done with \fastjet{3.0.3} \cite{Cacciari:2011ma} at the particle level, with no detector simulation.  The algorithms and groomers used in this study are available in the \texttt{Nsubjettiness} and \texttt{RecursiveTools} \fastjet{contrib}s \cite{fjcontrib}.

\section{Winner-Take-All at 100 TeV}
 \label{sec:pu}

High-energy jets are proxies for short-distance partons.  In the standard lore, the axis of a jet corresponds roughly to the direction of the initiating parton that subsequently showered and hadronized.  This is a useful picture when the jet's constituents arise primarily from final state radiation off the initiating parton.  However, at a hadron collider, there can be a significant jet contamination from (approximately) uncorrelated radiation---including underlying event, initial state radiation, and pileup---which are roughly uniformly distributed in pseudorapidity $\eta$.  Because the jet axis $\hat{a}$ is typically defined by the summed three-momenta of the jet's constituents,
\begin{equation}
\hat{a} \propto \sum_{i \in \text{jet}} \vec{p}_i,
\end{equation}
this uncorrelated radiation results in a significant displacement of the jet axis away from the initiating parton's direction.  This effect is referred to as recoil sensitivity \cite{Catani:1992jc,Dokshitzer:1998kz,Banfi:2004yd,Larkoski:2013eya}, and in order to be robust against jet contamination, we would like to define and use a recoil-insensitive axis.

The first recoil-free jet axis was introduced in \Refs{Georgi:1977sf,Thaler:2011gf,Larkoski:2014uqa}, where it was (eventually) called the ``broadening axis''.  The broadening axis $\hat{b}$ of a jet is defined by minimizing the value of broadening \cite{Rakow:1981qn,Ellis:1986ig,Catani:1992jc} with respect to it:
\begin{equation}
\hat{b} =  \argmin_{\hat{n}} \sum_{i \in \text{jet}} p_{Ti} \, R_{i\hat{n}},
\end{equation} 
where the sum runs over the particles in the jet, $p_{Ti}$ is the transverse momentum of particle $i$ with respect to the beam direction, and $R_{i\hat{n}}$ is the angle between particle $i$ and the axis $\hat{n}$.\footnote{In this language, the standard jet axis is given approximately by minimizing thrust \cite{Farhi:1977sg}, $\hat{a} \approx \argmin_{\hat{n}} \sum_{i \in \text{jet}} p_{Ti} R_{i\hat{n}}^2$.}  Unlike the standard jet axis definition, the minimization procedure for the broadening axis cannot be solved exactly, and numerical procedures to estimate $\hat{b}$ suffer from significant computational costs and spurious local minima.

  \begin{figure}[t]
 \includegraphics[height=6cm]{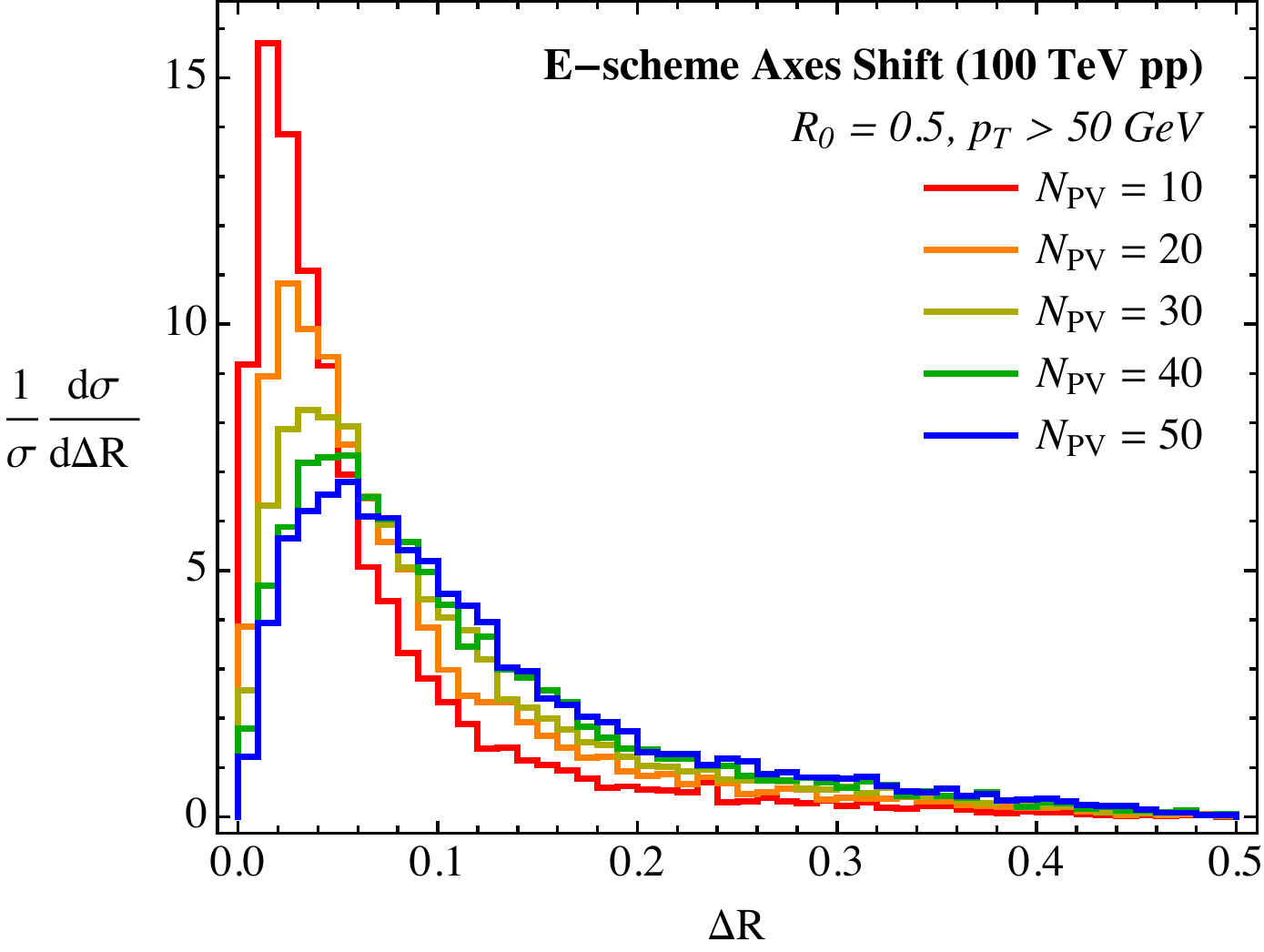}\\ \ \\
 \includegraphics[height=6cm]{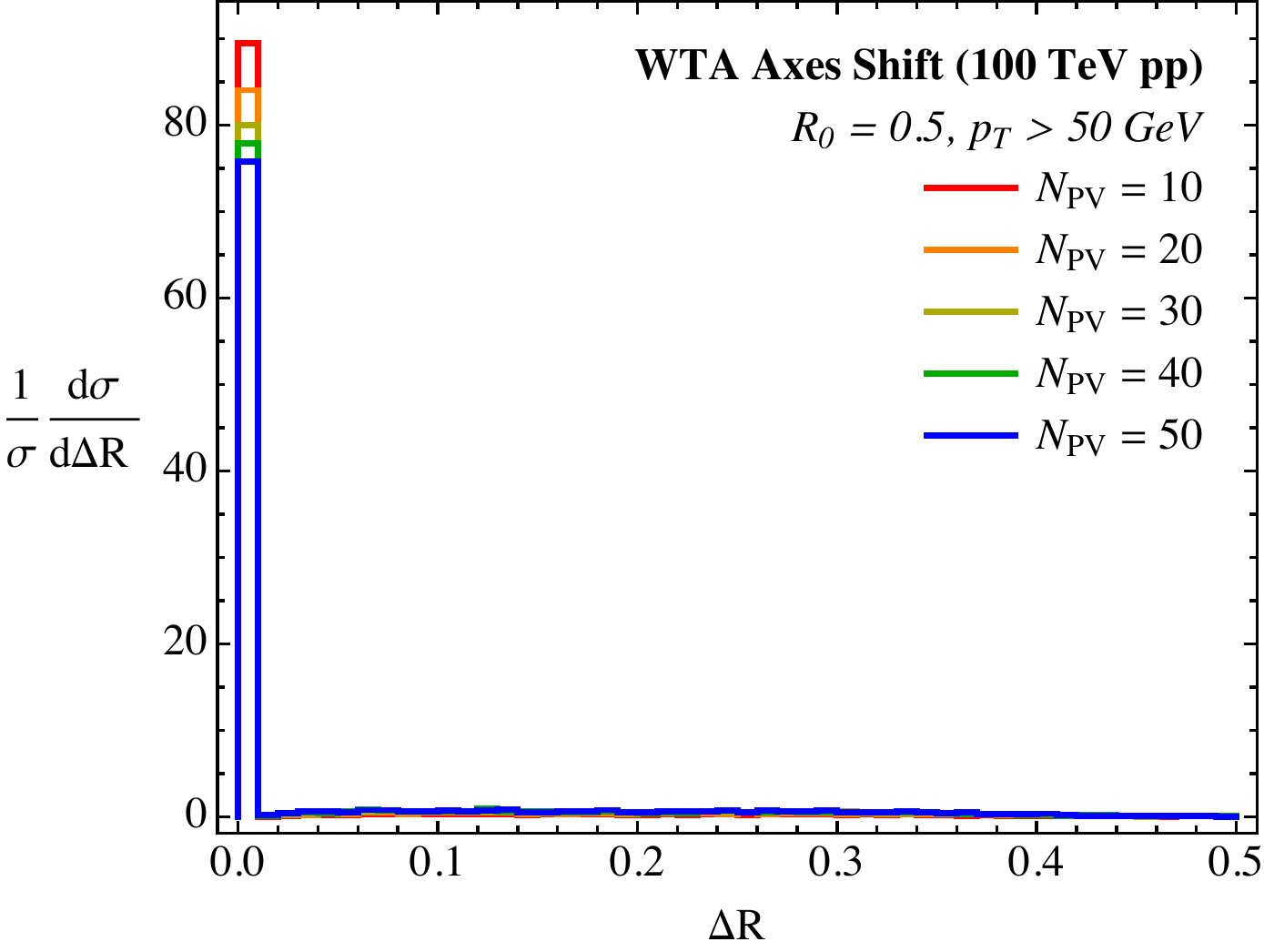}
 \caption{
The angular shift of the jet axis due to pileup, comparing the standard $E$-scheme jet axis (top) to the WTA jet axis (bottom), sweeping the number of pileup vertices $N_{\rm PV}$.
 \label{fig:axes_shift}}
 \end{figure}

A more practical recoil-free jet axis was presented in \Refs{Bertolini:2013iqa,Larkoski:2014uqa,Salambroadening}, where a ``winner-take-all (WTA) axis'' was defined by modifying a standard jet clustering algorithm.  Pairwise sequential jet algorithms are defined by two pieces:  a clustering metric and a recombination scheme.  The metric defines how close two particles are and whether they should be merged into a common jet.  Different metrics give rise to different jet algorithms---the \akt\ \cite{Cacciari:2008gp}, Cambridge/Aachen \cite{Dokshitzer:1997in,Wobisch:1998wt,Wobisch:2000dk}, and $k_T$ \cite{Catani:1993hr,Ellis:1993tq} algorithms are typical examples---and the only constraint on the metric is IRC safety.  The recombination scheme specifies how the momenta of two merging daughter particles should be mapped onto the momentum of the mother particle.  There is considerable freedom in this mapping (up to IRC safety).  The ubiquitous recombination scheme is the $E$-scheme \cite{Blazey:2000qt} where the daughter four-momenta are simply summed to define the mother momentum.  The $E$-scheme is manifestly sensitive to recoil because soft, wide-angle emissions in the jet will displace the mother from the harder of the daughter particles.

The WTA recombination scheme removes any effect from recoil.   In this scheme, the mother's $p_T$ is given by the scalar sum of the daughters' $p_T$, but the mother's direction is that of the harder daughter:
\begin{eqnarray*}
p_{TJ} &=& p_{Ti}+p_{Tj} ,\\
\phi_J&=&\begin{cases}
\phi_i, \quad p_{Ti}> p_{Tj} ,\\
\phi_j, \quad p_{Tj}> p_{Ti}  ,
\end{cases} \\
\eta_J&=&\begin{cases}
\eta_i, \quad p_{Ti}> p_{Tj} ,\\
\eta_j, \quad p_{Tj}> p_{Ti}  ,
\end{cases}
\end{eqnarray*}
where the daughters are $i,j$ and the mother is $J$.  This recombination scheme is IRC safe and is manifestly insensitive to the effects of recoil from soft, wide-angle emissions.  After running a pairwise jet algorithm with the WTA scheme (and any clustering metric), the final mother's direction defines a recoil-free jet axis.

To test the robustness of WTA jet axes, we generated dijet samples with $p_T>50$ GeV at a 100 TeV proton collider with various numbers of pileup vertices.\footnote{A 50 GeV jet at a 100 TeV collider---or even at Run II at the LHC---might seem a bit ridiculous, but here is used as a proof of concept.  The WTA axes are robust to recoil effects even for jets at very low $p_T$.}  The direction of the jets was compared before and after the addition of pileup, and the angle between the axes was computed.  Two jet algorithms were considered: \akt\ jets with standard $E$-scheme recombination and \akt\ jets with WTA recombination.  The jet radius in both samples is $R_0=0.5$.

In \Fig{fig:axes_shift}, we plot the angle $\Delta R$ between the jet axis before and after the addition of pileup.  As the number of pileup vertices $N_\text{PV}$ increases from 10 to 50, $\Delta R$ for the $E$-scheme axes increases noticeably, with a long tail extending to the jet radius of $0.5$.  By contrast, the WTA axes are amazingly rigid, and the vast majority of the jets have an identical axis ($\Delta R=0$) before and after the addition of pileup, even up to $N_\text{PV}=50$.  It should be stressed that for parton-level jets with $p_T= 50$ GeV, 50 pileup vertices at a 100 TeV collider corresponds to an ${\cal O}(1)$ increase in the $p_T$ of the jet, yet the WTA jet axes are still robust.

  \begin{figure}[t]
 \includegraphics[height=6cm]{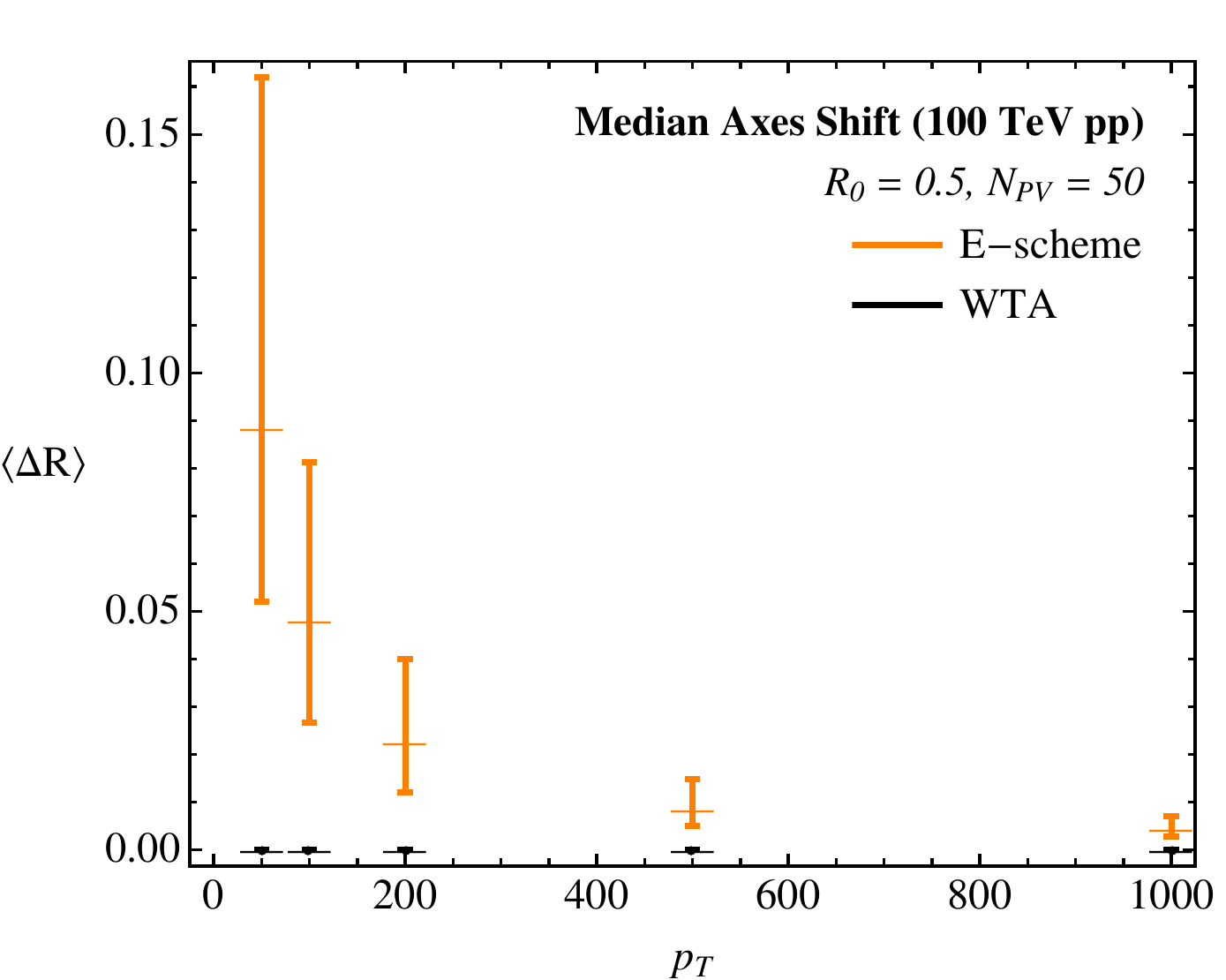}
 \caption{
The median angular shift due to pileup as a function of the jet $p_T$, comparing the $E$-scheme and WTA axes.  We have fixed $N_\text{PV} = 50$.  The lower (upper) error bar corresponds to the first (third) quartile.  For WTA, the median and the third quartile are in fact zero for the $p_T$ range studied.
 \label{fig:med_axes_shift}}
 \end{figure}

\Fig{fig:med_axes_shift} illustrates the $p_T$ dependence of the angle $\Delta R$ between the jet axis before and after the addition of pileup.  Here, the number of pileup vertices is fixed at $N_\text{PV} = 50$ and the $p_T$ of the jets ranges from $50$ to 1000 GeV.  We compare the median value of $\Delta R$ for the WTA and the $E$-scheme axes, with the lower and upper error bars corresponding to the first and third quartiles of the distribution.  As the $p_T$ of the jet increases, the median $\Delta R$ decreases for the $E$-scheme axes, but is still non-zero even at $p_T = 1000$ GeV.  By contrast, the median of the distribution of $\Delta R$ for the WTA axes is zero over the entire plotted range (as is the third quartile).  In \App{app:nonunipu}, we further demonstrate the robustness of the WTA axis to the effects of non-uniform pileup.
 
Beyond insensitivity to pileup, it is worth mentioning that recoil-insensitive axes are powerful for other reasons as well.  From a theoretical perspective, recoil-free observables are significantly easier to calculate in perturbative QCD than their recoil-sensitive counterparts \cite{Larkoski:2014uqa}.  Recoil-free observables also exhibit improved discrimination power between quark and gluon jets \cite{Larkoski:2013eya}.  From an experimental perspective, detector noise and finite resolution can fake the addition (or subtraction) of jet radiation, and a recoil-free axis is less sensitive to such effects.  Recoil-free axes can also be useful for validating pileup removal techniques, since the jet axis before and after the addition of pileup should be nearly identical. 

\section{Soft Drop at 100 TeV}
 \label{sec:sd}

While recoil-free observables offer some degree of robustness against pileup, one still needs techniques that actively identify and remove pileup contamination from a jet.  Such pileup removal procedures go under the general name of ``jet grooming'' and are vital for, say, accurately reproducing the mass of resonances that decay to jets.  Numerous grooming techniques have been developed, with filtering \cite{Butterworth:2008iy}, mass drop \cite{Butterworth:2008iy}, pruning \cite{Ellis:2009su,Ellis:2009me}, and trimming \cite{Krohn:2009th} being the most widely used at the LHC.  These procedures have been extensively validated on data \cite{Aad:2013gja,Chatrchyan:2013vbb} and their effects on jets and jet observables are well understood \cite{Walsh:2011fz,Dasgupta:2013ihk,Dasgupta:2013via}.  Here, we will review another jet grooming technique called ``soft drop declustering'' \cite{Larkoski:2014wba} and explain why it is well suited for 100 TeV jets.  Soft drop will also be important for the quasi-conformal observables we study in the next section.

To understand the motivation for soft drop, it is informative to review trimming \cite{Krohn:2009th} as an illustrative example of jet grooming.  To trim a jet, one first reclusters the jet's constituents using, e.g., the $k_T$ algorithm to form subjets of some small radius $R_\text{sub}$ (typically about $0.3$).  From the set of subjets, those whose $p_T$ fraction is less than some $\zcut$ (usually about $0.03$) are removed from the jet.  This has been shown to be an effective procedure for mitigating the effects of pileup on jet observables and allows for a more accurate reconstruction of the mass of boosted resonances.

Despite its successes, one key drawback of trimming is that it does not respect the approximate scale invariance of QCD (see \Ref{Dasgupta:2013ihk} for a discussion).   As the $p_T$ of a jet increases, jet radiation will move to smaller angular scales, such that no single $R_\text{sub}$ will be optimal over a wide range of $p_T$.  This is especially problematic for heavy objects like $W/Z$ or Higgs bosons, for which, at sufficiently large boosts, all of their decay products can lie within a cone of radius $R_\text{sub}$.  Of course, realistic detectors already have finite angular granularity, so a fixed $R_\text{sub}$ may not be too much of an issue in practice.  But given that jets at a 100 TeV collider can have such a large range of jet $p_T$ values, it is worth developing grooming procedures that do not have a fundamental angular limitation.

 \begin{figure}[t]
 \includegraphics[height=6cm]{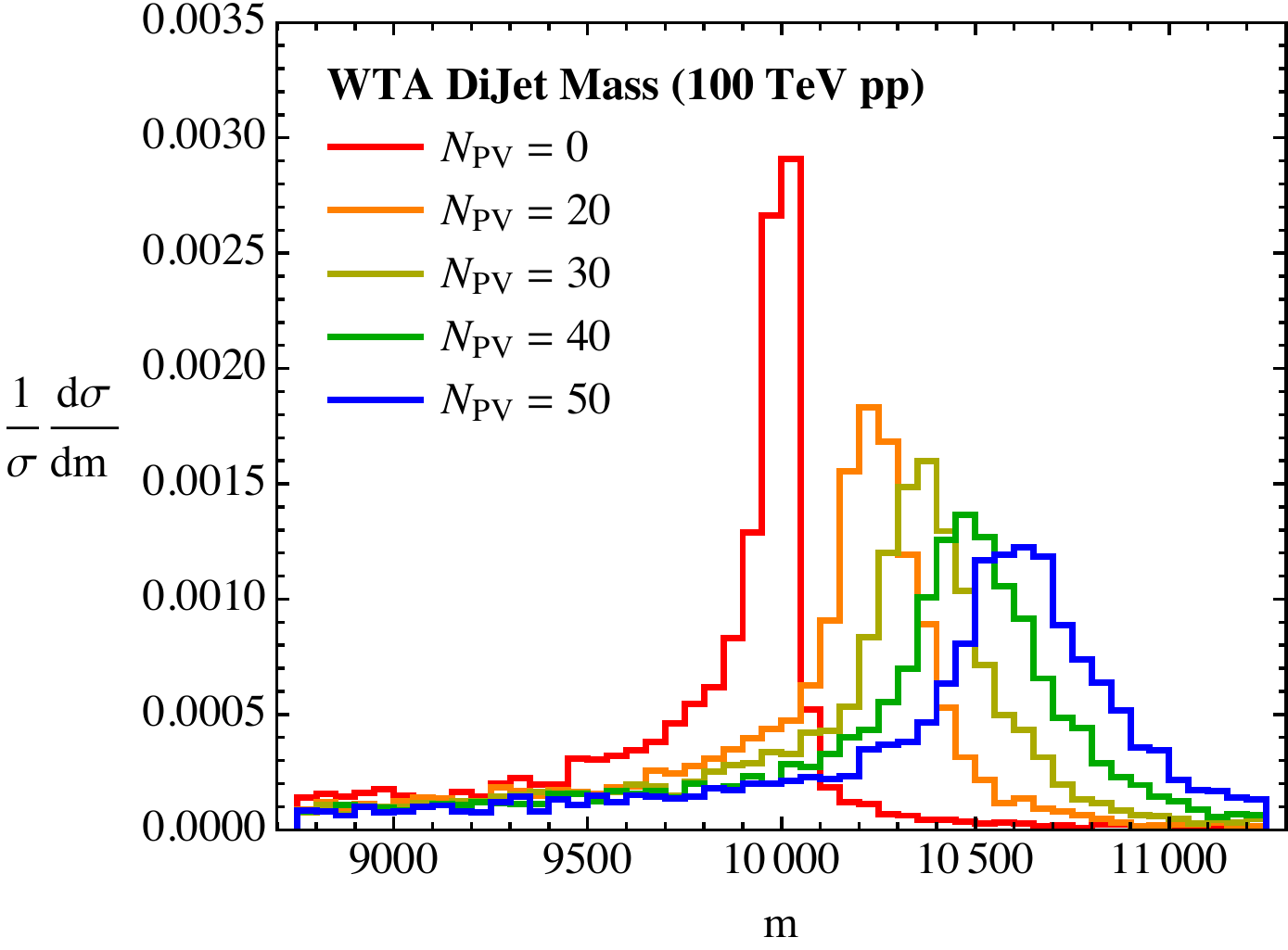}\\ \ \\
 \includegraphics[height=6cm]{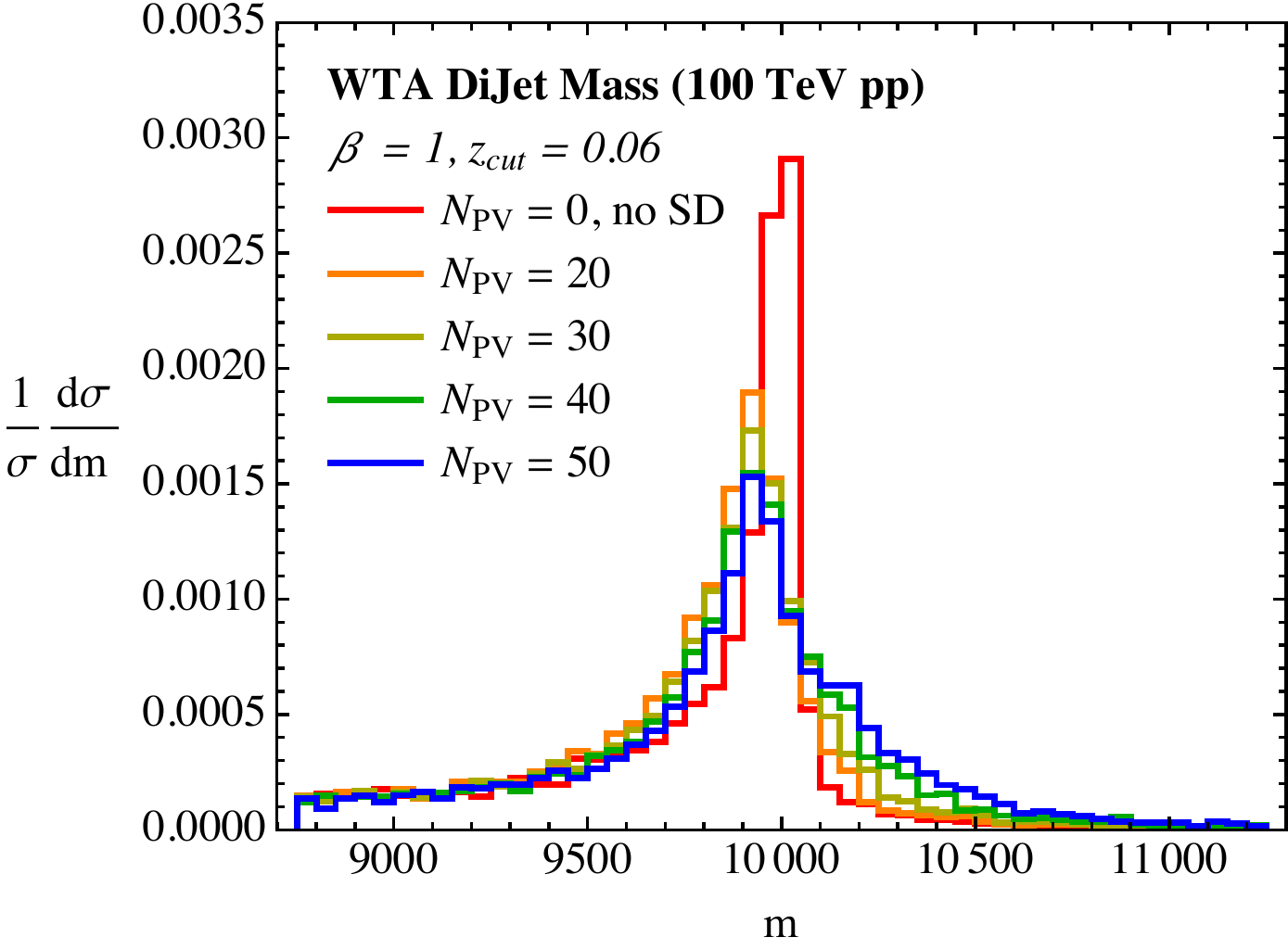}
 \caption{
Distribution of the reconstructed dijet invariant mass for a 10 TeV $Z'$ resonance, sweeping the number of pileup vertices $N_{\rm PV}$.  The top plot is without any pileup mitigation and the bottom plot is after the soft drop procedure.
 \label{fig:res_mass}}
 \end{figure}

\begin{figure}[t]
 \includegraphics[height=6cm]{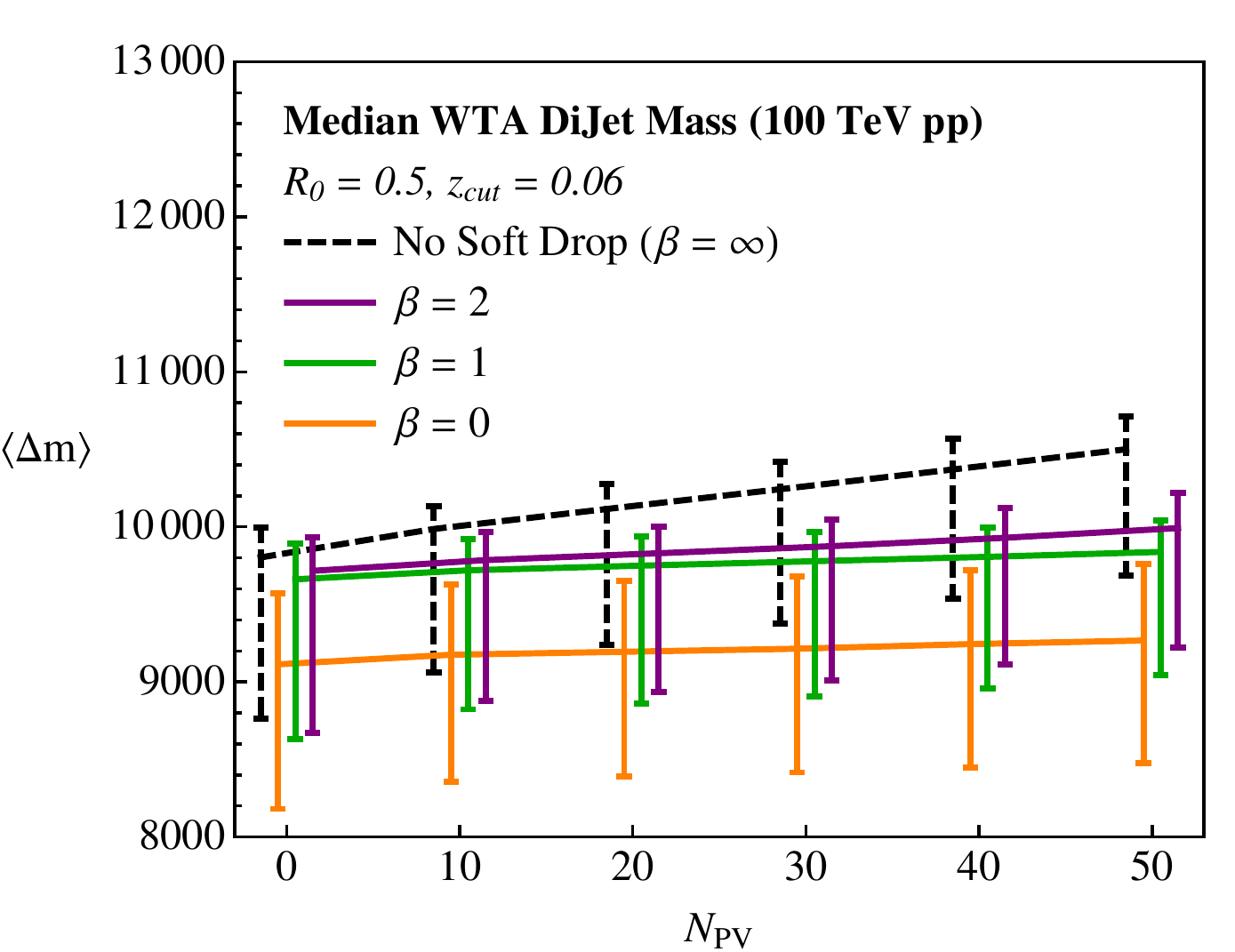}
 \caption{
Comparing the median of the dijet mass from $Z'$ decays with different values of the soft drop grooming parameter $\beta$ as a function of the number of pileup vertices $N_{\text{PV}}$.  The lower (upper) error bar corresponds to the first (third) quartile.    \label{fig:med_res_mass}}
 \end{figure}

The soft drop procedure is designed to dynamically identify the important angular range in a jet, and remove radiation outside of that range.  In that sense, soft drop behaves much like the (modified) mass drop procedure \cite{Butterworth:2008iy,Dasgupta:2013ihk}.  First, the jet is reclustered with the Cambridge/Aachen algorithm to construct an angular-ordered branching tree.  Starting at the trunk of the tree, if the first branching to pseudo-jets $i$ and $j$ fails the soft drop criteria:
\begin{equation}\label{eq:sd}
\frac{\min[p_{Ti},p_{Tj}]}{p_{Ti}+p_{Tj}} > \zcut \left(\frac{R_{ij}}{R_0}\right)^\beta  ,
\end{equation}
then the softer of the pseudo-jets is removed.  Here, $R_{ij}$ is the angle between the two pseudo-jets and $R_0$ is the jet radius.  The parameter $\zcut$ defines the threshold to remove soft radiation.  The angular exponent $\beta$ controls how aggressive the groomer is on different angular scales, with large positive $\beta$ corresponding to weak grooming and negative $\beta$ to very aggressive grooming.\footnote{Strictly speaking, negative $\beta$ is so aggressive that  it must be used as a tagger (instead of a groomer) in order to be IRC safe.  The distinction is that a groomer always returns a non-zero jet (even if it only has one constituent), whereas a tagger does not return a jet if \Eq{eq:sd} fails through the whole branching tree (i.e.~a jet with one remaining constituent is considered untagged).}  
This procedure continues up the tree, following the branch with the largest $p_T$, until it encounters a branching that satisfies \Eq{eq:sd}.  At this point the recursion terminates, leaving a jet with a groomed jet radius $R_g$.  Unlike trimming, the final angular scale $R_{g}$ is determined on a per-jet basis, such that soft drop respects the approximate scale invariance of QCD.  Note that the special case $\beta = 0$ corresponds to modified mass drop \cite{Butterworth:2008iy,Dasgupta:2013ihk}.\footnote{In the \texttt{RecursiveTools} \fastjet{contrib}, the $\beta = 0$ limit of soft drop is not quite identical to modified mass drop.  Soft drop defaults to grooming mode, whereas modified mass drop defaults to tagging mode.}

As a simple test of the efficacy of soft drop on pileup, we study the reconstruction of a massive dijet resonance.  We generated events of a 10 TeV narrow $Z'$ resonance that decays to light quarks at a 100 TeV collider with various numbers of pileup vertices.  The two hardest \akt\ jets are identified, using the WTA recombination scheme with radius $R_0=0.5$.  The upper plot of \Fig{fig:res_mass} shows the dijet invariant mass as a function of number of pileup vertices.  Not surprisingly, as the number of pileup vertices increases, the mass distribution both widens and drifts significantly.  The lower plot of \Fig{fig:res_mass} shows $m_{jj}$ after soft drop declustering with $\beta = 1$ and $\zcut = 0.06$.  As the number of pileup vertices increases, the average value of the groomed invariant mass is stable, and only slightly widens, nicely verifying the performance of soft drop.  

The soft drop parameters used in \Fig{fig:res_mass} were chosen because they resulted in the best reconstruction of the resonance mass.  To justify the choice of $\beta=1$, in \Fig{fig:med_res_mass} we plot the median dijet mass as a function of the number of pileup vertices for different values of $\beta$ (fixing $\zcut = 0.06$).  The upper (lower) error bars in \Fig{fig:med_res_mass} correspond to the first (third) quartile of the mass distribution.  From this figure, it is clear that $\beta = 0$ is too aggressive for this purpose, since the median value of the mass is decreased by several hundred GeV from baseline with no pileup.\footnote{Of course, one can make the $\beta = 0$ groomer less aggressive by decreasing $\zcut$, but we found one had to delicately adjust $\zcut$ to avoid over or under subtraction.  Having $\beta >0$ ensures that the hard core of the jet is never groomed away.}  Both $\beta = 1$ and $2$ both nicely mitigate the rise in the median mass without pileup, with $\beta = 1$ performing slightly better.  Also for $\beta = 1$, the locations of the first and third quartiles are relatively stable as pileup increases, comparable to the ungroomed mass distribution.

\section{Standard Candles at 100 TeV}
 \label{sec:sd_eloss}

Beyond the utility of soft drop as a jet grooming technique, soft drop can also be used to define new classes of jet observables with unique properties.  Several such observables were studied analytically in \Ref{Larkoski:2014wba}, including soft-dropped energy correlation functions \cite{Larkoski:2013eya,Banfi:2004yd} and the groomed jet radius $R_g$ itself. Here, we will study the $p_T$ fraction of a jet removed by soft drop.

Given an original jet with transverse momentum $p_{T0}$, the total fractional energy loss $\Delta_E$ is\footnote{For narrow jets, the distinction between energy fraction and transverse momentum fraction is negligible.}
\begin{equation}
\Delta_E = \frac{p_{T0}-p_{Tg}}{p_{T0}} ,
\end{equation}
where $p_{Tg}$ is the groomed jet transverse momentum.  This is the definition of the groomed jet energy loss originally given in \Ref{Larkoski:2014wba}.  Alternatively, we can measure the maximum energy fraction of the branches removed by soft drop:
\begin{equation}
z_{\max} = \max_{\text{failed branches}} \frac{p_{T,{\rm dropped}}}{p_{T, \rm{dropped}}+p_{T, {\rm kept}}}  .
\end{equation}
The reason for considering the two different definitions of the groomed $p_T$ loss will be explained below.  We will see that both $\Delta_E$ and $z_{\max}$ are quasi-conformal observables, in the sense that their distributions are remarkably insensitive to the energy scale of the jet.

In the case of pileup mitigation, neither $\Delta_E$ nor $z_{\max}$ are particularly interesting, since they correspond to radiation that one wants to remove from a jet since it likely comes from contamination.  In the absence of pileup, though, $\Delta_E$ and $z_{\max}$ are sensitive probes of the intrinsic soft radiation captured in a jet, so one could imagine measuring them to calibrate the response of a detector to soft (perturbative) physics.  

For $\beta > 0$, $\Delta_E$ and $z_{\max}$ are IRC safe, so the distributions can be computed in perturbative QCD.  The calculation of the distribution of $\Delta_E$ was presented in detail in \Ref{Larkoski:2014wba}, including the resummation of large logarithms.  To leading-logarithmic accuracy with fixed coupling $\alpha_s$, the cumulative distribution was found to be
\begin{align}
\label{eq:deltaEdist}
\Sigma(\Delta_E) &= 
\frac{\log \zcut - B_i}{\log \Delta_E - B_i}\nonumber\\
&\quad+\frac{\pi \beta}{2 C_i \alpha_s} 
\frac{1-e^{-2\frac{\alpha_s}{\pi}\frac{C_i}{\beta}\log\frac{\zcut}{\Delta_E}\left(\log \frac{1}{\Delta_E} + B_i\right)}}{(\log\Delta_E - B_i)^2},
\end{align}
where $C_i$ is the color factor of the jet ($C_F = 4/3$ for quarks, $C_A = 3$ for gluons) and $B_i$ are the subleading terms in the splitting functions  ($B_q = -3/4$ for quarks, $B_g = -\frac{11}{12}+\frac{n_f}{6C_A} $ for gluons, where $n_f$ is the number of light flavors).  Of course, this distribution can be improved with running coupling effects, higher order resummation, and nonperturbative corrections.  To the accuracy that we will work in this paper, the distribution for $z_{\max}$ is identical to \Eq{eq:deltaEdist}, with the replacement $\Delta_E\to z_{\max}$.

However, there is an important distinction between $\Delta_E$ and $z_{\max}$.  Because $\Delta_E$ is determined by the sum total of all radiation that was groomed away, this observable depends on multiple emissions within the jet.  The effect of multiple emissions is not included in \Eq{eq:deltaEdist}, as it is formally beyond the accuracy of that expression.  On the other hand, $z_{\max}$ is defined by a single groomed branch, and so the effects of multiple emissions are minimal.  Therefore, we expect that \Eq{eq:deltaEdist} will give a better description of $z_{\max}$ than $\Delta_E$.

The form of \Eq{eq:deltaEdist} is not particularly enlightening, but various expansions can be taken to illuminate its behavior.  First, we can expand in $\alpha_s$, which results in
\begin{equation}
\Sigma(\Delta_E) = 1-\frac{\alpha_s}{\pi}\frac{C_i}{\beta}\log^2\frac{\zcut}{\Delta_E} + {\cal O}\left(\left(\frac{\alpha_s}{\beta}\right)^2\right)  ,
\end{equation}
which, for $\beta>0$, is a Taylor series in $\alpha_s$, illustrating its IRC safety.  However, as $\beta\to0$, every term in this expansion diverges, meaning that the soft-dropped energy loss is not IRC safe for $\beta = 0$.  Nevertheless, from the full resummed expression we can take the $\beta\to 0$ limit first, which produces
\begin{equation}\label{eq:b0el}
\Sigma(\Delta_E)_{\beta = 0} = \frac{\log \zcut - B_i}{\log \Delta_E - B_i},
\end{equation}
restricted to $\Delta_E<\zcut$.  As mentioned above, $\beta = 0$ corresponds to the (modified) mass drop procedure \cite{Butterworth:2008iy,Dasgupta:2013ihk}.  That $\Delta_E$ for $\beta=0$ is IRC unsafe but still calculable when all-orders effects are included means that it is a Sudakov-safe observable \cite{Larkoski:2013paa}.

 \begin{figure}[t]
 \includegraphics[height=6cm]{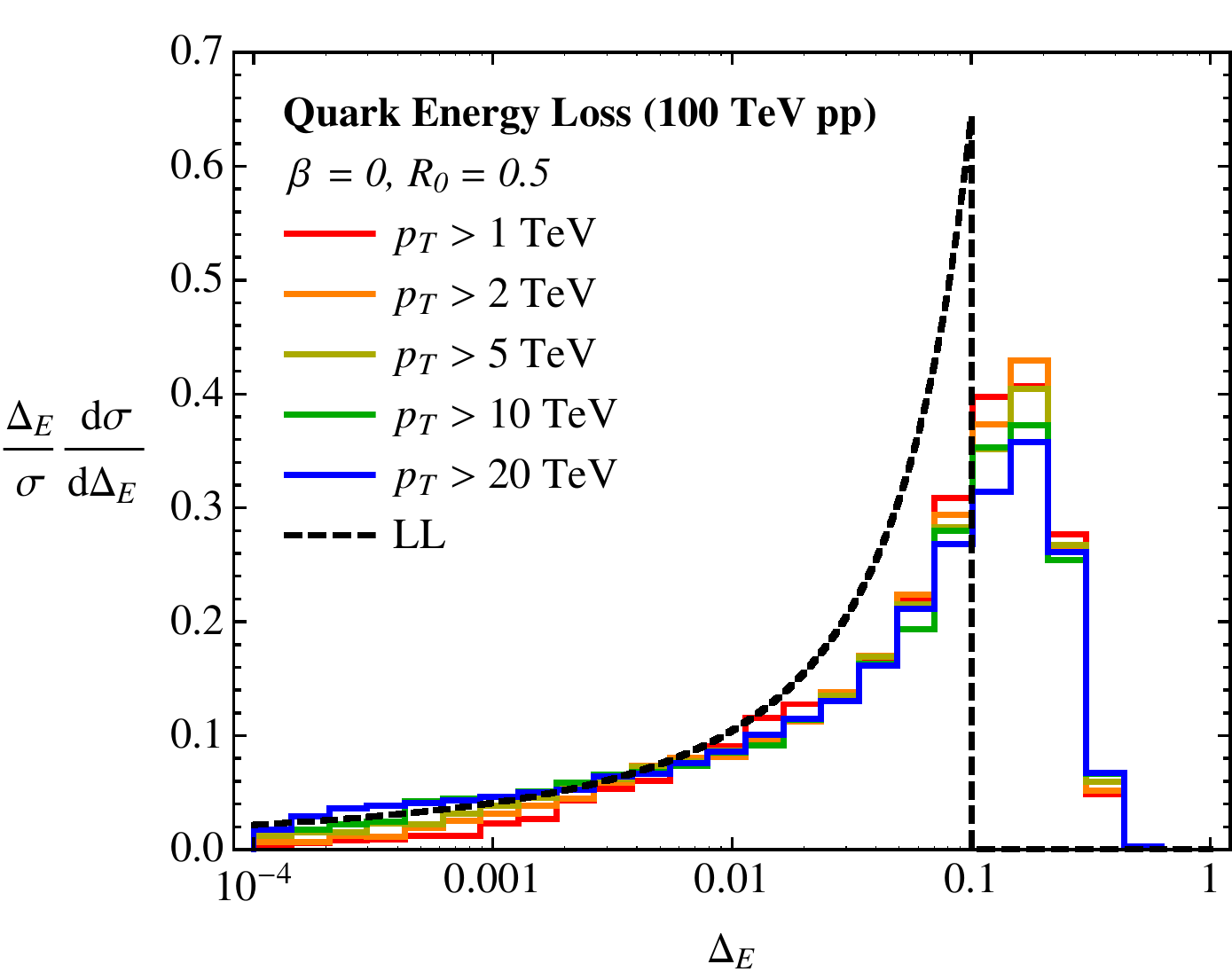}\\ \ \\
 \includegraphics[height=6cm]{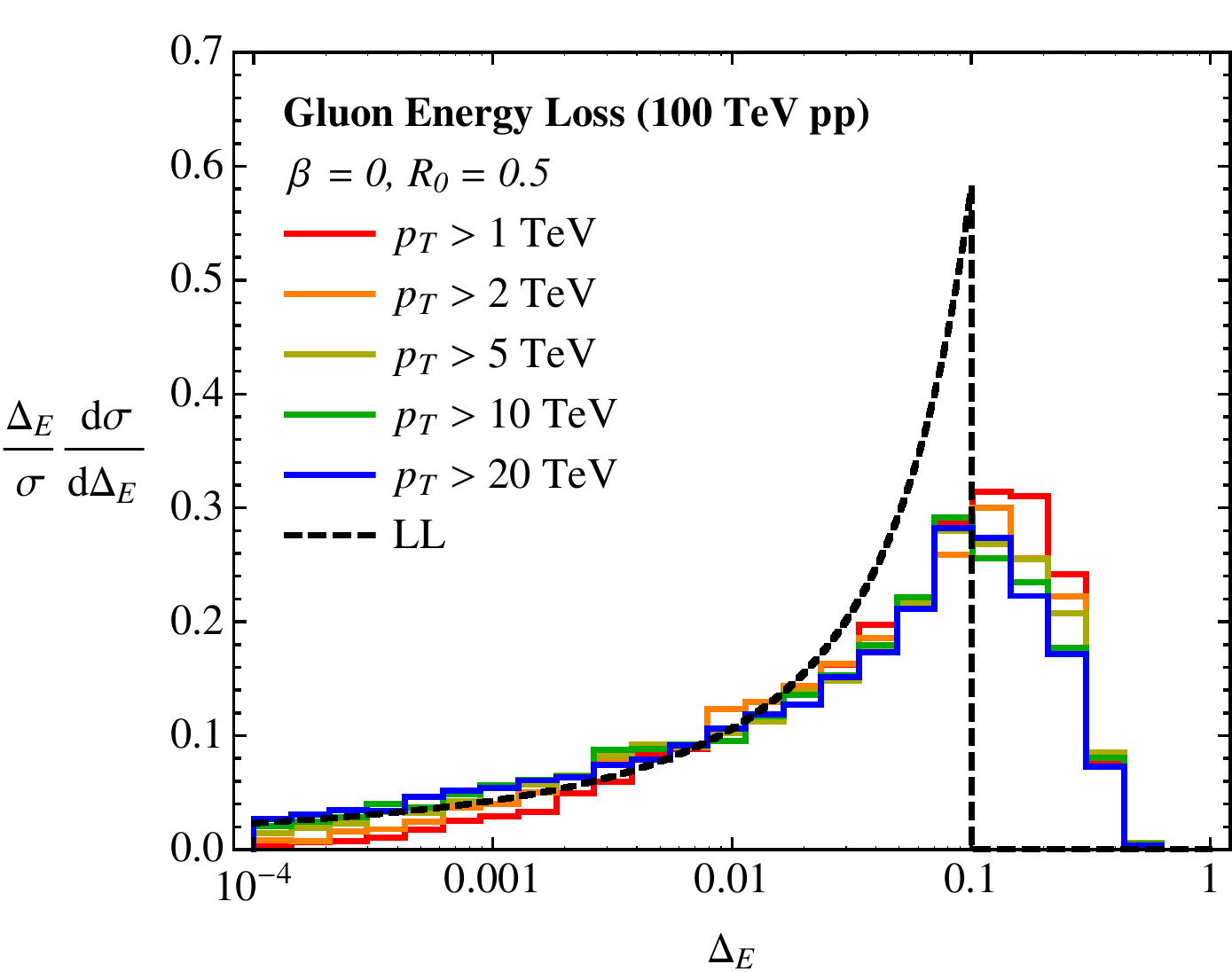}
 \caption{
Distribution of the total fractional energy loss $\Delta_E$ after soft drop for quark jet (top) and gluon jets (bottom) over a range of $p_T$ values.  ``LL'' is the distribution computed from \Eq{eq:b0el}, with the appropriate $B_i$ factors for quark and gluon jets.
 \label{fig:e_loss}}
 \end{figure}
 
  \begin{figure}[t]
 \includegraphics[height=6cm]{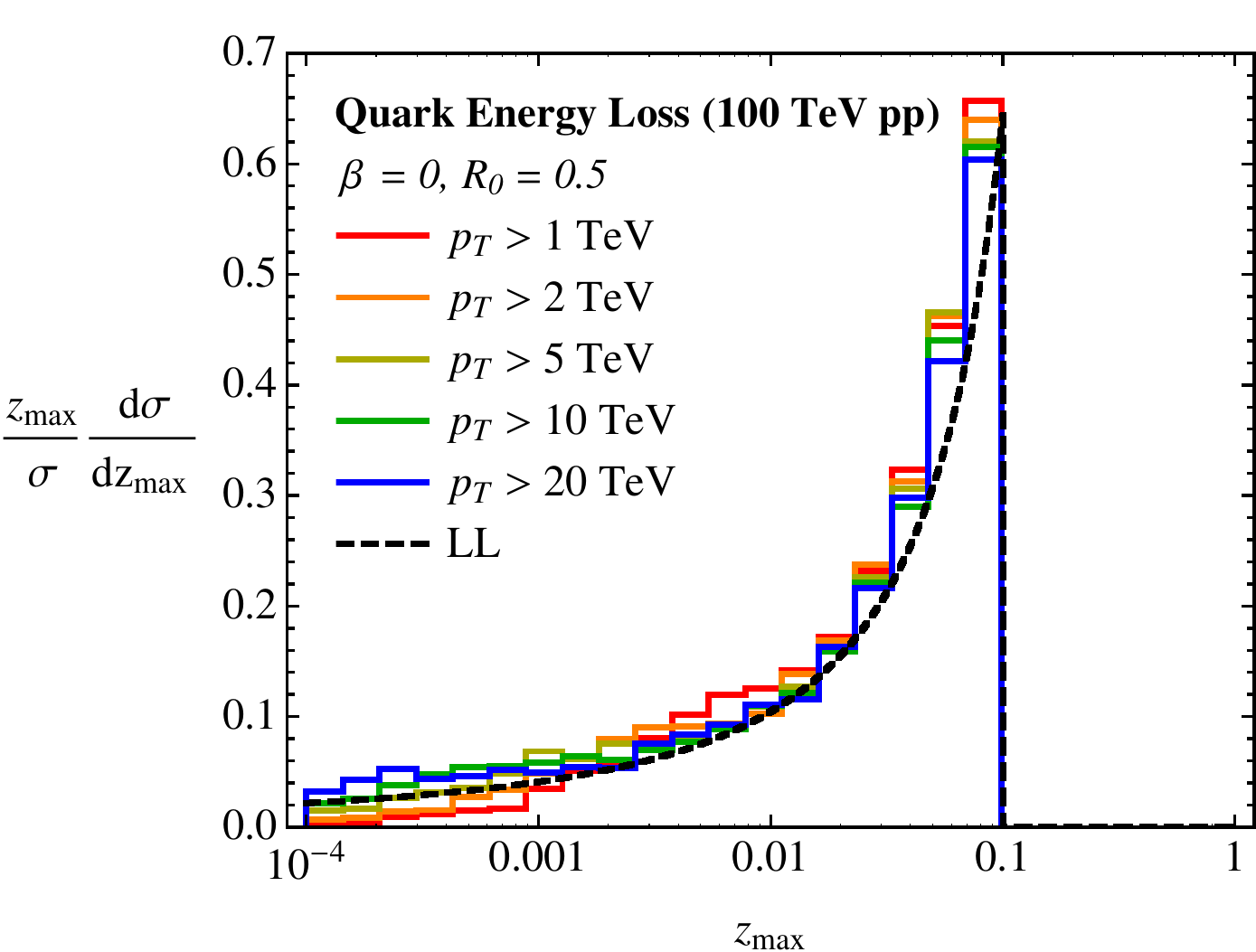}\\ \ \\
 \includegraphics[height=6cm]{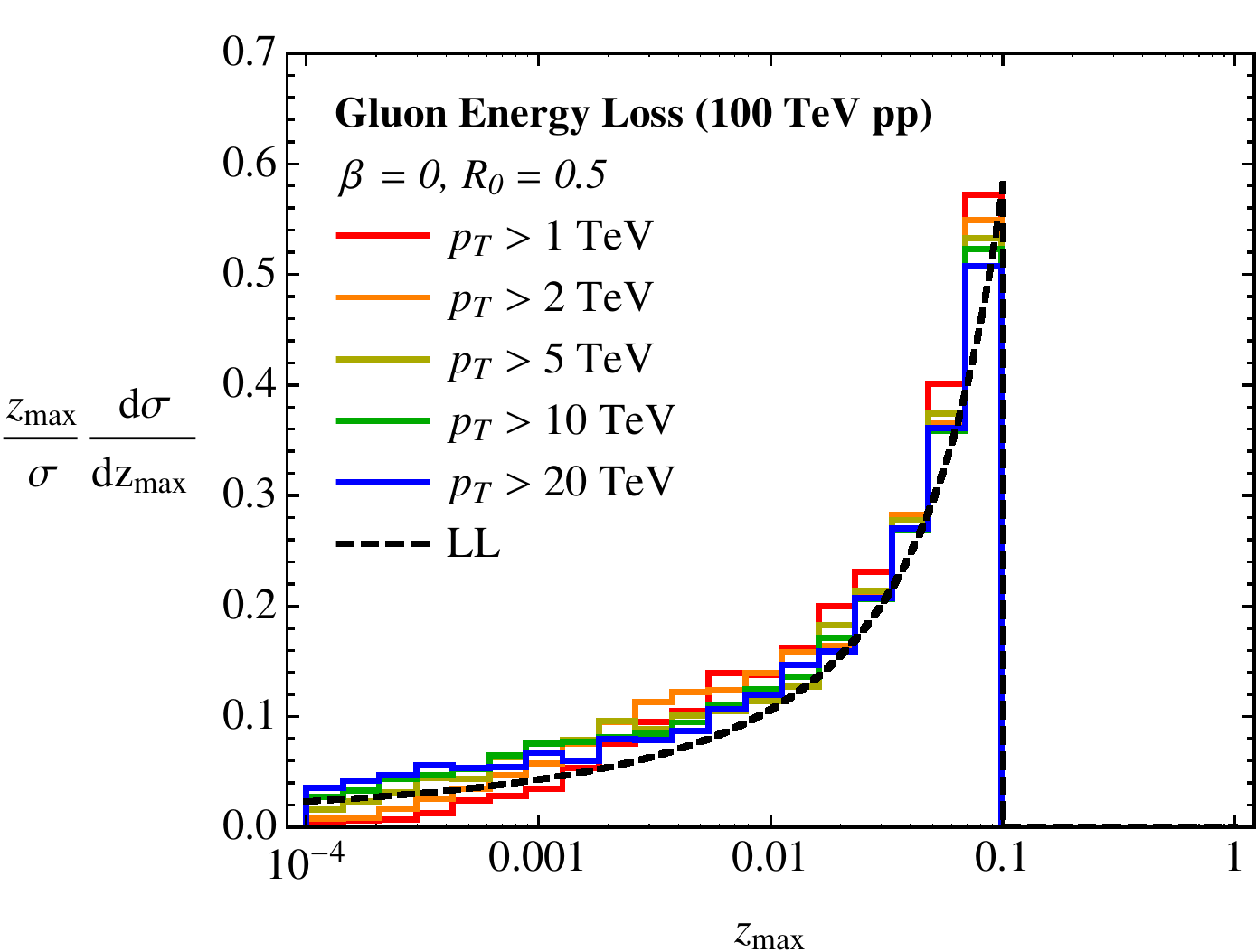}
 \caption{
Same as \Fig{fig:e_loss}, but for the maximum fractional energy loss $z_{\max}$.
 \label{fig:e_loss_max}}
 \end{figure}

The $\beta = 0$ distribution in \Eq{eq:b0el} is fascinating.  In the fixed-coupling limit, it is independent of $\alpha_s$.  This implies that the distribution is only weakly dependent on the energy scale of the jet (i.e.~it is quasi-conformal), with all dependence suppressed by the (small) $\beta$-function of QCD.  It is also independent of the total color of the jet, and so the distribution should be nearly identical for quark and gluon jets, with the dependence on the flavor of the jet entering from the subleading $B_i$ terms.  This illustrates some of the surprising features of Sudakov-safe observables: because their distributions are not required to be a Taylor series in $\alpha_s$, they can have peculiar dependence on the coupling.  

We can use parton shower simulations to test the degree to which $\Delta_E$ is independent of the jet scale and jet flavor.  We generated dijet events over a range of transverse momenta at a 100 TeV collider.  Unlike the previous sections, we do not include pileup in this analysis, so as to isolate the physics of the soft drop procedure on the perturbative radiation in the jet.  We plot the distribution of $\Delta_E$ in \Fig{fig:e_loss} and $z_{\max}$ in \Fig{fig:e_loss_max} on \akt\ jets with radius $R_0=0.5$ for pure quark and gluon jet samples.\footnote{The quark jets come from $qq\to qq$ and the gluon jets from $gg\to gg$.  We ignore any subtleties regarding sample dependence or the precise definition of jet flavor.}  The $p_T$ of the jets ranges from 1 TeV to 20 TeV, and the distributions at different $p_T$'s lie on top of one another, until very small values of $\Delta_E$ or $z_{\max}$, where honest non-perturbative effects dominate.  Also, the quark and gluon distributions are remarkably similar, except for large values of $\Delta_E$ where subleading perturbative effects are important.  On these plots, we have also included the calculated distribution from \Eq{eq:b0el}, appropriate for quark or gluon jets.  Especially for $z_{\max}$, the leading-logarithmic prediction nicely matches the parton shower.  As expected from the discussion of multiple emissions above, \Eq{eq:b0el} gives a better prediction for the distribution of $z_{\max}$ than $\Delta_E$, especially for values near $\zcut$.

\section{Conclusions}

Jets provide a unique probe into the dynamics of physics in (and beyond) the standard model.  As particle physics experiments move to ever higher energies and luminosities, we have the opportunity to rethink standard approaches to jets.   By incorporating new concepts like recoil insensitivity and Sudakov safety, we have the potential to increase both theoretical and experimental control over jet observables.

In this paper, we have highlighted three ways that jet analyses could be improved at a 100 TeV proton collider.  The WTA recombination scheme allows the definition of a recoil-free jet axis, which improves the robustness of jet identification in the presence of pileup.  The soft drop declustering procedure respects the approximate scale invariance of QCD, allowing it to groom away jet contamination over a wide dynamical range.  Sudakov-safe observables are not constrained to be a Taylor series in $\alpha_s$, allowing for interesting probes of QCD that are not possible with standard IRC-safe observables, including quasi-conformal and quasi-flavor-blind observables.  While the focus of this paper was on a 100 TeV collider, these same techniques are of course relevant at the LHC, since there are similar pileup issues for 14 TeV proton collisions, and jet studies in heavy-ion collisions could benefit from increased robustness to the QCD fireball.

The most provocative proposal in this paper is using Sudakov-safe observables as a standard candle for jets.  Indeed, actually measuring the $\Delta_E$ or $z_{\max}$ distributions is quite challenging experimentally, since it requires a detailed understanding the energy composition of a jet.  For a 10 TeV jet, measuring $\Delta_E$ down to $0.001$ requires understanding 10 GeV substructure.  We suspect that the scale- and flavor-independence of these observables is not unique, however, and we look forward to developing Sudakov-safe observables that are more tractable experimentally.  That said, this standard candle does offer an interesting and ambitious target for designing 100 TeV detectors.

\begin{acknowledgments}
This paper is based on a talk by A.L.~at the ``Workshop on Physics at a 100 TeV Collider'' at SLAC.  We thank the organizers and participants of this workshop for valuable discussions, and especially Michael Peskin for encouragement.  We thank Simone Marzani, Duff Neill, and Gregory Soyez for collaborating on the work cited here.  We also thank Gavin Salam, Gregory Soyez, and T.J.~Wilkason for help developing the \fastjet{contrib}s used in these studies.  This work is supported by the U.S. Department of Energy (DOE) under cooperative research agreement DE-FG02-05ER-41360.  J.T.~is also supported by the DOE Early Career research program DE-FG02-11ER-41741 and by a Sloan Research Fellowship from the Alfred P. Sloan Foundation.
\end{acknowledgments}

\appendix

\section{Effect of Non-Uniform Pileup}
\label{app:nonunipu}

 \begin{figure}[t]
 \includegraphics[height=6cm]{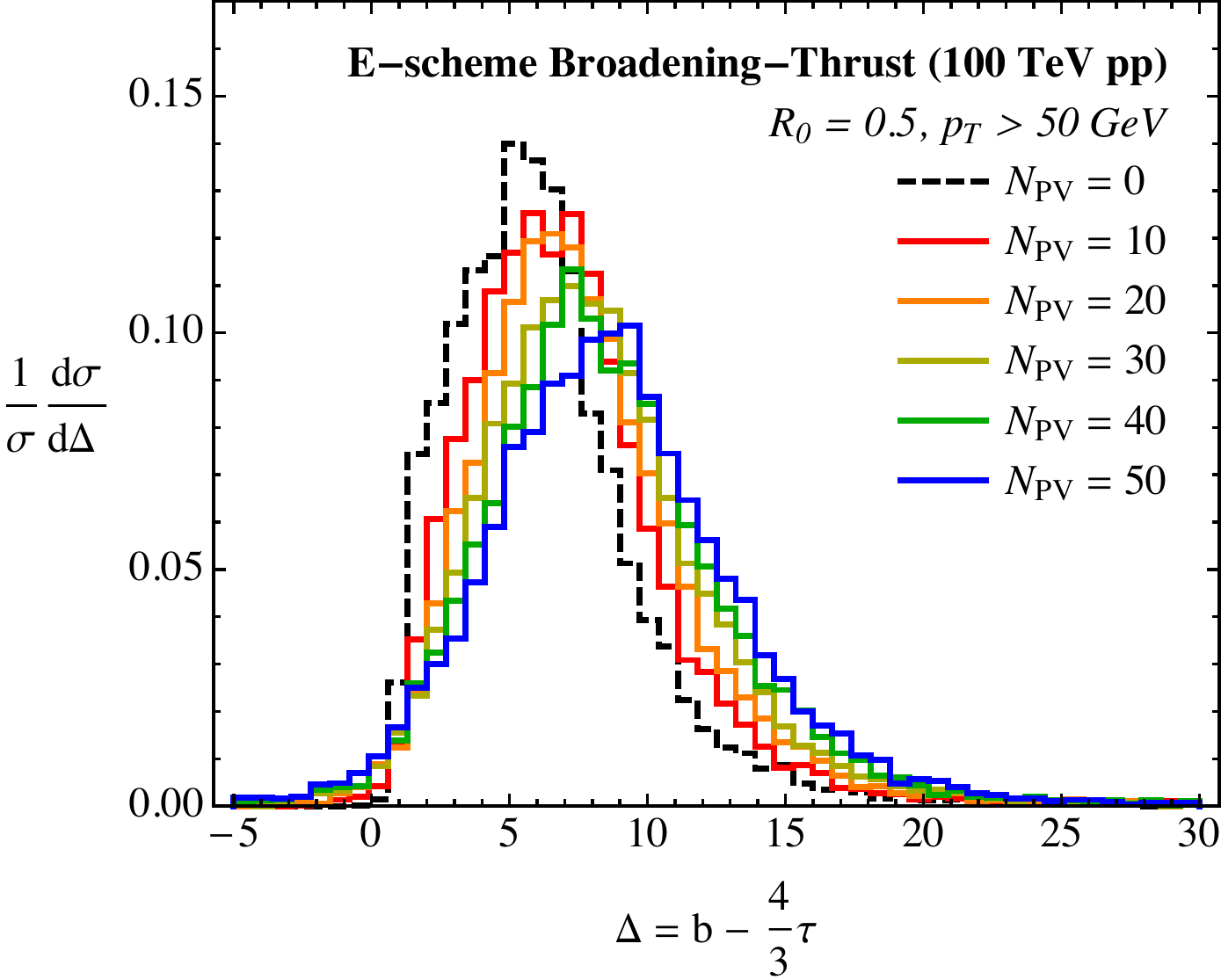}\\ \ \\
 \includegraphics[height=6cm]{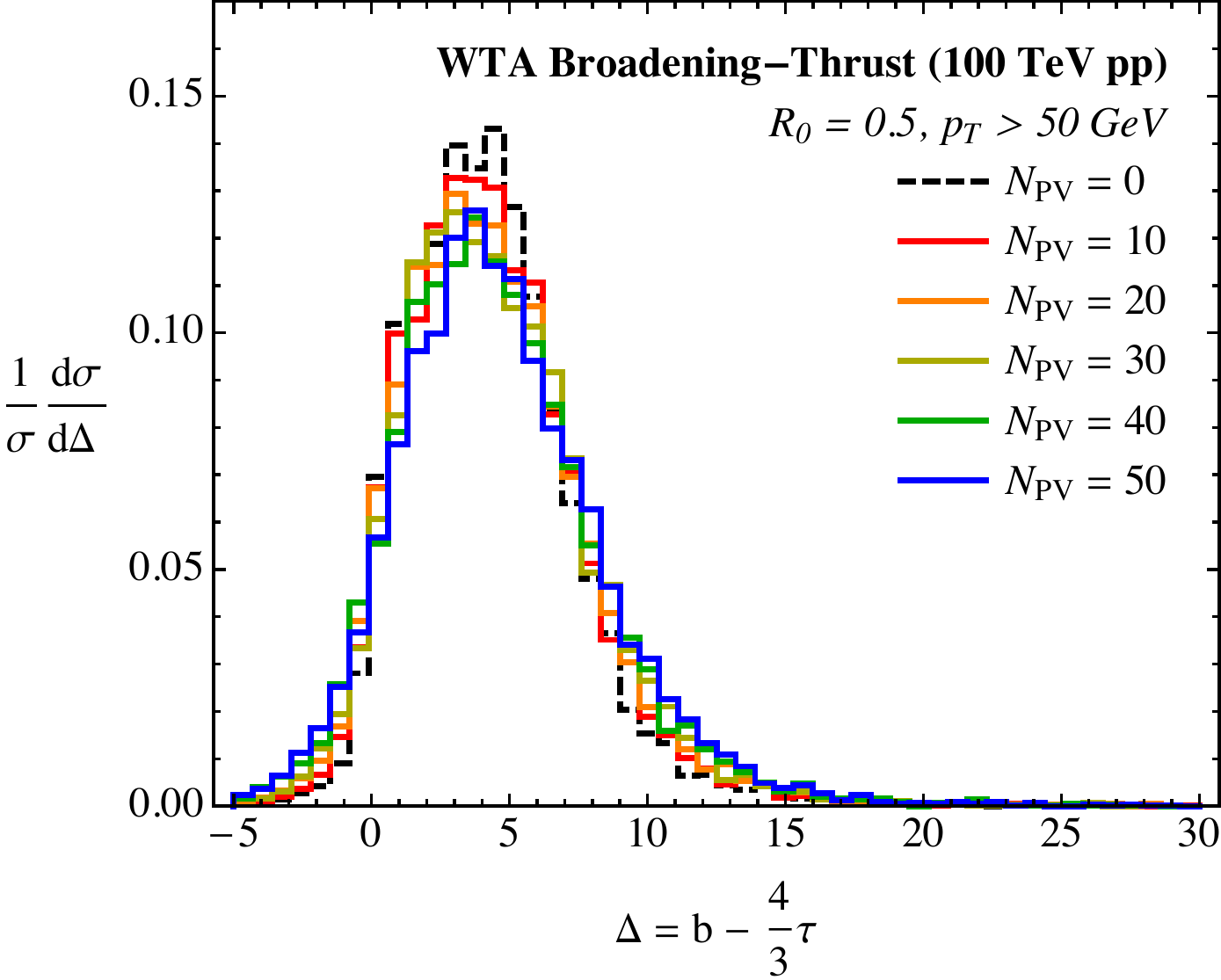}
 \caption{The distributions for $\Delta = b - \frac{4}{3}\tau$ (i.e.\ broadening minus scaled thrust), comparing the standard $E$-scheme jet axis (top) to the WTA axis (bottom), sweeping the number of pileup vertices $N_{\rm PV}$.   The $4/3$ scaling factor is chosen such that $\Delta$ is insensitive to uniform jet contamination, so this is a direct test of non-uniform pileup dependence.
 \label{fig:broad_thrust}}
 \end{figure}

We saw in \Sec{sec:pu} that the WTA axis was robust to pileup, but one might wonder if the same robustness could be achieved through jet area subtraction \cite{Cacciari:2007fd,Cacciari:2008gn,Soyez:2012hv}.  Crucially, the effect of recoil cannot be removed through jet area subtraction alone, since that technique assumes  pileup contamination is uniformly distributed over the jet.  While the jet energy can be largely corrected using a uniform subtraction, the axis shift seen in \Fig{fig:axes_shift} really corresponds to the response of the jet axis to non-uniformities in the pileup.  

To test the effect of non-uniform pileup more directly, we can study observables that, at least on average, are independent of uniform radiation in the jet.  First, consider the dimensionful angularities $e_\alpha$ \cite{Berger:2003iw,Almeida:2008yp,Ellis:2010rwa} measured about some axis $\hat{n}$:
\begin{equation}
e_\alpha = \sum_{i\in\text{jet}} p_{Ti}\left(
\frac{R_{i\hat{n}}}{R_0}
\right)^\alpha,
\end{equation}
where $R_{i\hat{n}}$ is the angle between particle $i$ and $\hat{n}$ and the angular exponent $\alpha>0$ for IRC safety.  Different angularities have a different sensitivity to pileup because of the different angular weighting.  We can exploit this fact to define an observable that is on average insensitive to uniform contamination, by taking an appropriate linear combination of two angularities $e_\alpha$ and $e_\beta$.  For contamination with a fixed $p_T$, the correct linear combination is found from ($\theta \equiv R_{i\hat{n}}/R_0$)
\begin{eqnarray}
\langle e_\beta - x e_\alpha\rangle &\propto& \int_0^1 \theta d\theta \left(\theta^\beta - x\theta^\alpha\right) = 0\nonumber \\
&&  \Rightarrow x = \frac{\alpha+2}{\beta+2}  ,
\end{eqnarray}
where $\theta d\theta$ is the angular measure for uniform radiation in the jet.  Therefore, the observable
\begin{equation}
\Delta = e_\beta - \frac{\alpha+2}{\beta+2}e_\alpha  ,
\end{equation}
is insensitive to uniform contamination on average.  For concreteness, we will consider the difference between jet broadening $b$ ($\beta = 1$) and jet thrust $\tau$ ($\alpha = 2$),
\begin{equation}
\Delta = b - \frac{4}{3}\tau. 
\end{equation}

In \Fig{fig:broad_thrust}, we plot $\Delta$ for the same sample of jets as in \Fig{fig:axes_shift}, comparing angularities measured with respect to the $E$-scheme axis and the WTA axis.  For both axis choices, the distributions broaden somewhat as the number of pileup vertices increases.  However, while the recoil-sensitive $E$-scheme axis exhibits a significant drift in the average value of $\Delta$, the recoil-free WTA axis gives a distribution that remains centered at the same $\langle \Delta \rangle$ value.   Thus, the recoil-free jet axis is more robust to non-uniform radiation in the jet than the standard recoil-sensitive jet axis.

\bibliography{100tev}

\end{document}